# Domain Mapping for Volumetric Parameterization using Harmonic Functions

Vikash Gupta[1], Hari K. Voruganti[2] and Bhaskar Dasgupta[3]

[1]Asclepios Team Project, INRIA Sophia Antipolis Mediterranee, France. vikash.gupta@inria.fr
[2]Dept. of Mech. Engg., National Institute of Technology, Warangal, India, harikumar@nitw.ac.in
[3]Dept. of Mech. Engg., Indian Institute of Technology, Kanpur, India. dasgupta@iitk.ac.in

**ABSTRACT**

Volumetric parameterization problem refers to parameterization of both the interior and boundary of a 3D model. It is a much harder problem compared to surface parameterization where a parametric representation is worked out only for the boundary of a 3D model (which is a surface). Volumetric parameterization is typically helpful in solving complicated geometric problems pertaining to shape matching, morphing, path planning of robots, and isogeometric analysis etc. A novel method is proposed in which a volume parameterization is developed by mapping a general non-convex (genus-0) domain to its topologically equivalent convex domain. In order to achieve a continuous and bijective mapping of a domain, first we use the harmonic function to establish a potential field over the domain. The gradients of the potential values are used to track the streamlines which originate from the boundary and converge to a single point, referred to as the shape center. Each streamline approaches the shape center at a unique polar angle $(\theta)$ and an azimuthal angle $(\psi)$. Once all the three parameters (potential value $\phi$, polar angle $\theta$, azimuthal angle $\psi$) necessary to represent any point in the given domain are available, the domain is said to be parameterized. Using our method, given a 3D non-convex domain, we can parameterize the surface as well as the interior of the domain. The proposed method is implemented and the algorithm is tested on many standard cases to demonstrate the effectiveness.

## 1    INTRODUCTION

With the advent of powerful 3D scanning techniques, there is an abundance of 3D shape data available to the research community involved in geometric analysis of shapes. The next important issue is: how to mathematically represent the shapes and use them for various applications such as morphing, texture mapping, shape matching and remeshing etc.? For all these applications, a continuous parameterization of the complete model is necessary. Volumetric parameterization provides a continuous representation of both the boundary and interior of the domain. This is accomplished by establishing a bijective mapping between the given model and a convex domain. Volumetric parameterization is vital for many applications involving the physics of the model. For example, structural analysis, heat transfer and fluid flow analysis, electromagnetic etc. A recent approach called isogeometric analysis for accurate analysis involves developing and exact geometric model using parameterization techniques.

Though several methods have been developed for parameterization of surfaces in 3D, not many methods are proposed for parameterization of the entire volume of a 3D model. Most of the surface parameterization methods cannot be extended to volume. Not even twenty articles exist in the literature on volumetric parameterization. In the following section, surface parameterization techniques are reviewed briefly followed by the existing method for volume parameterization.

## 1.1 Surface Parameterization

The surface parameterization domain is thoroughly explored and a summary of the published works can be found in a set of review articles [19,4,22,8]. Many methods for surface parameterization involve partitioning the surface into simpler patches using cuts. Such methods suffer from inherent problems because of the continuity issues along cuts and their junctions [20]. A spherical parameterization method was introduced by Praun and Hoppe [18] which directly parameterizes the entire domain to a spherical domain by minimizing stretch based measure. This method was extended by Asirvatham et al [2] so that multiple objects can be parameterized simultaneously without losing any properties of mapping. Brechbuhler et al. [3] used parameterization as a tool for shape description of 3D objects. Posing it as an optimization problem, they expanded the parameterized surfaces into spherical harmonics and proposed methods to parameterize the surfaces which are invariant to translation, rotation or scaling. Mapping a genus-0 mesh onto a spherical surface by minimizing discrete harmonic energy is the most common technique used in spherical parameterization. A spherical parameterization method targeting surface fitting was developed by Li et al [14] where parameterization is obtained by minimizing the discrete harmonic energy. Friedel et al. [5] used a triangle energy minimization technique to establish spherical parameterization for a given domain.

Though all these methods provide surface parameterization with some variations in the energy terms involved, distortion of the domain, computational efficiency and applicability to applications etc, none of them are directly applicable to the case of parameterization of an entire volume. Hence there is a need for development of such methods. Some of the attempts to tackle this more difficult problem are summarized below.

## 1.2 Volume Parameterization

Unlike surface parameterization, volume parameterization involves parameterization of both boundary and interior. In 3D, the boundary of the domain is a surface. Hence surface parameterization can be interpreted as a subproblem of volume parameterization. In fact, there are some methods like Patro et al [17] which develop parameterization for the surface in the first stage and use it as a boundary condition for obtaining volume parameterization. Such methods are suitable both for surface and volume parameterization. The drawback of such methods is that they involve partitioning the boundary and hence associated issues mentioned in 1.1.

Another group of methods uses the theory of harmonic functions to establish a potential field over a given domain. Harmonic functions are a good choice of potential functions because of their elegant properties including maximum principle and mean-value property. Li et al [13] developed a harmonic volume mapping method. Given a boundary mapping they develop volume mapping using fundamental solution method. The quality of mapping depends on the given boundary mapping. Wang et al [27] developed two techniques which focus on computer graphics and medical imaging. This method also involves boundary mapping and then volume mapping using heat flow method. But such mapping is difficult when the boundary surface is highly convoluted.

A mapping between two reference free-form models was established by using volumetric parameterization by Wang et al. [26] while keeping the spatial relationship between the two models intact. Very recently, isogeometric analysis has become an important area of application for volume mapping which may accelerate research in volume mapping. Martin et al [15] developed yet another method based on harmonic function and subsequently designed an algorithm for B-spline modeling of the given model. Such a model is directly useful in isogeometric analysis. A method called *Cubecover* involving user intervention is proposed by Nieser et al [16] for parameterization of 3D volumes with cubes which is similar to the *Quadcover* for surfaces [9].

In this paper, we present a novel approach for the problem of volumetric parameterization of genus-0 3D regions. We use harmonic functions to establish a uniform potential field across the domain. Then the streamlines are tracked using the potential gradient within the domain. Due to the property of the streamlines, they approach the shape center at unique angles. When combined with the computed potential of the internal points, the three parameters establish volumetric parameterization. Our approach is analogous to a heat conduction phenomenon. Our method ensures bijective mapping of complex non-convex shapes and it is demonstrated using several typical shapes such as a star-fish, the human face and biomolecules.

The paper explains the principle and required mathematical background for volumetric parameterization. In Section 2, we present the definitions pertaining to convex and non-convex domains. In section 3, a detailed analogy between the volumetric parameterization problem and the heat transfer phenomenon is drawn. The harmonic function and its associated properties are discussed in Section 4, which is followed by a step-by-step algorithm for volumetric parameterization in Section 5. Finally the results are presented in Section 6 and concluding remarks in Section 7.

## 2   CONVEX AND NON-CONVEX DOMAINS

A convex domain is one in which a straight line joining any two points in the domain is completely contained within it. The interior of a circle is an example of a convex domain in two dimensions because, given any two points inside the circle, they can be joined by a straight line, no part of which lies outside the circle. In three dimensions, the above example can be extended to a sphere. Contrary to convex domains, a shape in which a straight line segment connecting two arbitrary points contained in it need not lie completely inside, is called non-convex.

The motivation behind mapping a non-convex domain is to be able to uniquely parameterize the interior and boundary points of the domain. For example, different points on a circle can be parameterized using the polar angles. In figure 1(a), the radial lines OA, OB and OC make different angles with the horizontal OX. In a similar manner, parameterizing any convex domain is a trivial problem. But if we follow the same procedure to parameterize a non-convex domain, by considering a some point within the domain as center, the problem of non-uniqueness emerges, which means that two or more points have the same set of parameters. Unlike the convex case, radial lines OA and OB corresponding to two distinct points A and B make same angle with horizontal OX. Thus, the above mentioned problem of non-uniqueness is illustrated. To address this problem, a method is developed with the objective of representing the non-convex domains also in a way similar to a sphere i.e. every point of the domain is expressed by three unique parameter values. Once this is achieved, a domain is said to be parameterized. To achieve this, straight radial lines are replaces by curved lines and concentric spheres by potential shells. The proposed method is originated from the observations of the properties of potential field and streamlines in the analysis of heat transfer problems.

## 3   DOMAIN MAPPING AND HEAT TRANSFER: ANALOGY

The need for unique mapping has led us to look into heat transfer mechanisms. We can draw an analogy between the heat transfer problem and the one pertaining to volumetric parameterization of non-convex domains. Figure 2 shows the heat flow-lines (streamlines) and the isothermal contours for a 2D non-convex shape. A constant temperature heat sink is located at the shape center and the boundary is maintained at a constant elevated temperature. Under these conditions, a temperature gradient is set up between the boundary and the heat sink at the shape center, which decreases as the boundary is approached. It should be noted that the lines emanating from the boundary and approaching the shape center (streamlines) never intersect each other and approach the shape center at a *unique* angle. This angle value is assigned to all the points of a streamline (i.e. angle value is constant along a streamline), while the temperature decreases from the boundary to the shape center along a streamline. Thus, we have two parameters (temperature of a point and the angle subtended by the streamline at the centre) which can be used to locate any point within the domain.

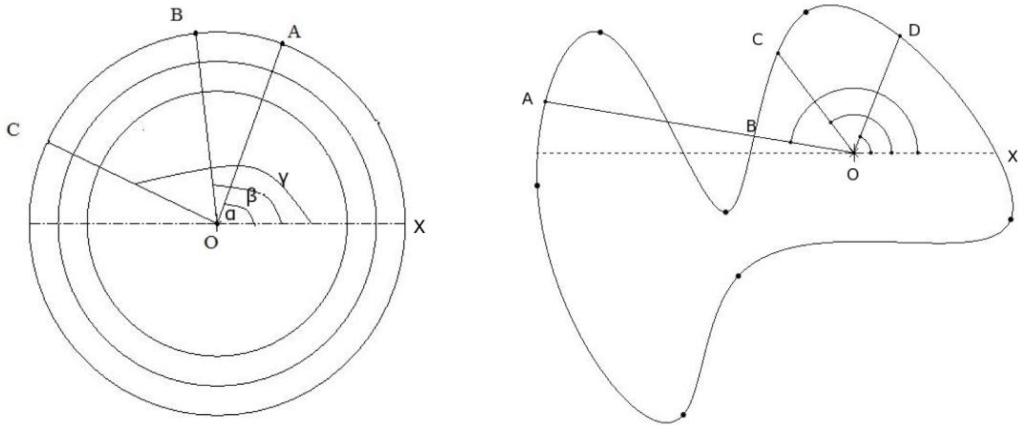

Fig.1: Boundary points subtend different angles at the shape center O for convex domains, while two boundary points (A and B) subtend the same angle at the shape center in a non-convex case. (a) Convex domain: Circle, and (b) Non-convex domain.

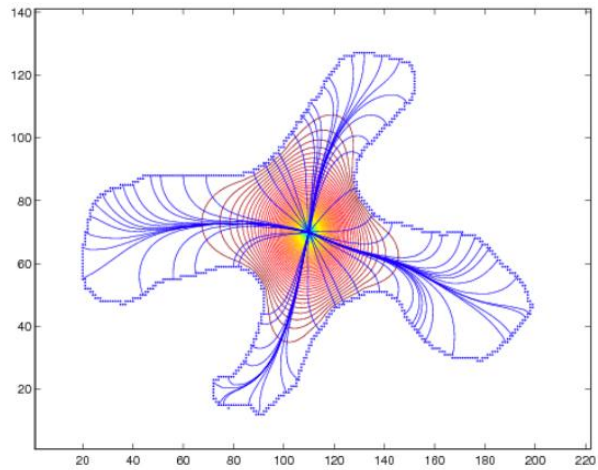

Fig.2: Streamlines in a non-convex domain

The above analogy can be extended to any non-convex domain in three dimensions, the only difference being that, in three dimensions, a pair of angles (polar and azimuthal angles, similar to a spherical coordinate system) is required to completely define the angle of approach. Therefore, the two angles along with the temperature value parameterize a 3D domain.

Identifying this fact, that solving a boundary value problem to establish a potential field over the domain directly helps in parameterizing the given domain. We use harmonic functions to establish the required field.

## 4  HARMONIC FUNCTIONS

Harmonic functions play a key role in establishing bijective mapping in our method. A twice differentiable real-valued function $\phi: U \to R$, where $U \subseteq R^n$ is some domain, is called harmonic if its **Laplacian** vanishes over $U$, i.e. if $\nabla^2 \phi = 0$. Mathematically, a function satisfying the Laplace equation,

$$\nabla^2 \phi \equiv \sum_{i=1}^{n} \frac{\partial^2 \phi}{\partial x_i^2} = 0, \tag{1}$$

Where $x_i$ is the $i$-th Cartesian coordinate and $n$ is the dimension of the domain under study, is called a harmonic function. The mean value and the maximum value property are the ones which make harmonic functions useful for many applications [6,11].

## 4.1   Mean-Value Property

If a sphere $B(x,r)$ with center $x$ and radius $r$, is completely contained in the domain $U$, then the value $\phi(x)$ of the harmonic function at the center of the ball is given by the average of values of $\phi$ on the surface of the sphere. Interestingly, this average value is also equal to the average of the values of $\phi$ inside the sphere.

## 4.2   Maximum Principle

If $\phi$ is a harmonic function, then according to the maximum principle, $\phi$ cannot have local extrema within the domain $U$. By definition of harmonic functions, their Laplacian should be zero. For a local extremum to exist the second order partial derivatives of the function should have the same sign. If all of them have the same sign, their sum will never be zero and thus they will never be able to satisfy Laplace's equation.

## 5   METHODOLOGY

The description of the domain or model is the only input required. A description can be a set of points, surface mesh or tetrahedral mesh. Depending on the input data, an appropriate pre-processing is needed. When input data is in the form of a surface mesh, the internal mesh is generated using TetGen [24]. Once the internal mesh is generated, equipotential shells are created within the domain using harmonic functions. Then the streamlines are tracked, which intersect the equipotential shells orthogonally. The various steps involved in the algorithm discussed in this paper can be outlined as follows.

1. Generate the tetrahedral mesh from triangular surface mesh using TetGen.
2. Assign the internal points generated to a 3-dimensional grid. This process is called discretization.
3. Separate the external, internal and boundary points by flagging them and choose the shape center.
4. Apply the boundary conditions.
5. Compute potential of all the internal points, using harmonic functions.
6. Compute streamlines.
7. Calculate the polar $(\theta)$ and azimuthal $(\psi)$ angles corresponding to the streamlines. These angles are the parameters.
8. Parameterization is completed through interpolation over these values.
9. Inverse mapping by interpolation.

The angles computed in step 7 are plotted on a graph. This kind of plot is called the atlas and gives a visual estimate of the distribution of the boundary points on the parameterized $(\theta - \psi)$ domain. These angles combined with the potential values of interior points give the volumetric parameterization of the domain.

## 5.1 Pre-processing

A wide range of 3D shapes are available on the internet at various repositories [1,12]. These models are only triangulated surface meshes and contain no information about the internal meshing or nodes. But, we need to compute the potential of the interior points. The internal mesh information is generated using TetGen [24]. Depending on the range of the coordinates of the domain, a 3-dimensional grid is constructed. The points within the domain are assigned to the different grid nodes. This operation is called *discretization* and it helps in further potential computation. The grid nodes are classified as exterior, boundary and interior points based on neighborhood information.

An appropriate choice of the **shape center** is very important for all computations. The most important criterion for choosing a shape center is that it should be located well within the domain so that it is easily reachable by all or most of the boundary points. A better location of shape center facilitates better and more accurate potential distribution within the domain which leads to a faster and more accurate computation of the streamlines.

## 5.2 Boundary Conditions

After the data is discretized as mentioned above, we apply the *Dirichlet boundary conditions* for the boundary and shape center i.e. we assign a potential value of 1 ($\phi = 1$) to all the boundary nodes and a potential value of 0 ($\phi = 0$) to the shape center. During the entire computation (iterations) the value of $\phi$ is unaltered. All other interior points are assigned random potential values as initial values. In addition, we also assign a potential of $1+\epsilon$ to the points just outside the boundary, the next layer of external points is assigned a potential of $1+2\epsilon$ and so on until we have four layers, where $\epsilon$ is a very small number of the order of $10e-4$ or less. We will see later in the paper that this kind of external potential assignment prevents the streamlines from digressing outside the domain of interest.

## 5.3 Potential Computation

The iterative finite difference method is used to solve Laplace's equations. If $f(x,y)$ is a harmonic function, its second derivative as derived using Taylor series expansion and neglecting the higher order terms:

$$\frac{\partial^2 f}{\partial x^2} = \frac{f(x_{i+1}, y_i, z_i) - 2f(x_i, y_i, z_i) + f(x_{i-1}, y_i, z_i)}{h^2}$$

$$\frac{\partial^2 f}{\partial y^2} = \frac{f(x_i, y_{i+1}, z_i) - 2f(x_i, y_i, z_i) + f(x_i, y_{i-1}, z_i)}{k^2}$$

$$\frac{\partial^2 f}{\partial z^2} = \frac{f(x_i, y_i, z_{i+1}) - 2f(x_i, y_i, z_i) + f(x_i, y_i, z_{i-1})}{l^2} \quad (2)$$

where $h$, $k$ and $l$ are the step-sizes in the $x$, $y$ and $z$ directions, respectively. We choose equal step-sizes in all three directions for simplicity. The potential $\phi$ is used as a candidate for harmonic functions. Thus, using Laplace's equation, we get

$$\phi(x_i, y_j, z_k) = \frac{\phi(x_{i+1}, y_j, z_i) + \phi(x_{i-1}, y_i, z_i)}{6h^2} + \frac{\phi(x_i, y_{i+1}, z_i) + \phi(x_i, y_{i-1}, z_i)}{6h^2} + \frac{\phi(x_i, y_i, z_{i+1}) + \phi(x_i, y_i, z_{i-1})}{6h^2}. \quad (3)$$

The above potential values are computed iteratively until the maximum difference between two successive computations on any of the nodes is less than a pre-defined value $\zeta$, which we call the tolerance. Thus, if $\phi_j$ and $\phi_{j+1}$ are the two values of potential ($\phi$) computed in the $j$-th and $(j+1)$-th iterations, the termination criterion for the computation will be

$$max \mid \phi_{j+1} - \phi_j \mid < \zeta. \quad (4)$$

A wise choice of $\zeta$ is important. A very small value of $\zeta$ will increase the accuracy of potential computation, thus enabling good streamline tracking, but it will need huge computational resources.

But a wisely chosen value of $\zeta$ will result in quite acceptable accuracy fairly fast. Typically the value of $\zeta$ ranges from $10^{-3}$ for shapes as simple as, sphere to $10^{-6}$ for complicated shapes.

### 5.4 Streamlines

Streamlines are flow lines or gradient lines. They are orthogonal to potential shells. Thus, they can be characterized with $\nabla \phi$. A streamline starts from the boundary and proceeds towards the shape center. Further, because of the inherent property of the streamlines, they intersect with the equipotential surfaces within the domain at right angles. In the case of a spherical shape, the equipotential surfaces are just different concentric spheres within the domain, the shape center being the same as the geometric center of the sphere, the radial lines will be the streamlines and they evidently intersect the concentric spheres orthogonally. In the case of a sphere, they are straight lines, but in the case of non-convex domains, they may curve in order to satisfy the orthogonality criterion.

Ideally we want the streamlines emerging from the boundary nodes to end at the shape center. But as the streamlines reach well within the domain, where the irregularities from the boundary are smoothened out, we can terminate the computation, because the terminating points of the streamlines form a convex boundary around the shape center which is sufficient to obtain unique angles. The streamline tracking problem is essentially equivalent to solving an ordinary differential equation. If $X(t) = [x, y, z]^T$ is a coordinate vector, then the differential equation for streamlines is,

$$\dot{X}(t) = -\eta \nabla \phi [X(t)] \quad (5)$$

where $\eta$ is called the normalization parameter. We have used the adaptive *Runge-kutta* method for solving these equations. Using adaptive step-size was essential as it reduces the computational cost. The streamlines take large steps and tend to converge quickly in simple regions of the domain whereas they take appropriately small steps in the complicated regions. The ordinary differential equation (ODE) solver requires the potential for the non-grid points within the domain, i.e., for the intermediate points. This is achieved by effecting a bilinear fitting locally. Thus

$$\phi(x, y, z) = p_1 xyz + p_2 xy + p_3 yz + p_4 zx + p_5 x + p_6 y + p_7 z + p_8 \quad (6)$$

gives us the potential value for any point within the domain. It uses the potential values of the eight neighboring grid-nodes to frame the equations. This set of eight equations is solved for $p_i$'s. using the Gaussian elimination method. Thus, the potential at any random point is evaluated.

### 5.5 Mapping the Domain: Volumetric Parameterization

After the streamlines are tracked, the end-points of the streamlines approach the shape center with unique sets of angles (polar ($\theta$) and azimuthal ($\psi$) angles). The end points of the streamlines are obtained in the Cartesian coordinate system. We apply the Cartesian to spherical transformation to obtain these angles ($\theta$ and $\psi$) desired for the mapping as,

$$\psi = atan2(y, x), \theta = atan2(\sqrt{x^2 + y^2}, z).$$

Since the genus-0 domains are topologically equivalent to a sphere, once these angles are available, we can use the set of parameters ($r, \theta, \psi$) to create a sphere of unit radius.

$$x' = r \sin \theta \cos \psi, y' = r \sin \theta \sin \psi, z' = r \cos \theta;$$

where $r = 1$ for unit sphere. In other words, the given domain is mapped to a sphere. Once these mapped coordinates are available, the bijectivity of the mapping can be demonstrated using the atlas plots in the next step as mentioned earlier in Section 5.

## 6 RESULTS AND DISCUSSION

In this section we present the parameterized models. The 3D models are obtained from the various scanning repositories. The entire algorithm was implemented using the C programming language. In the following sections, we present the results obtained after testing our algorithm on several complicated models. It can be seen that there is no bijectivity loss in the mapping of a domain, which establishes the algorithm.

### 6.1 Case 1: Synthetic Domain

A synthetic domain with 9597 vertices and 30073 triangles is shown in figure 3. It is a non-convex domain but not very complicated. It has been chosen for demonstration because the mapping can be seen easily. A parameterization has been developed for this model by following the procedure explained in this paper. As can be seen from the atlas (figure 4), the mapping is bijective. The mapping can also be demonstrated by plotting the mapped domain i.e., sphere as shown in figure 5. Most of the published papers show the mapping results using such plots though it is not the best way because one cannot see the entire map. The atlas plot is a better way to show the mapping which we have adapted for the rest of the paper.

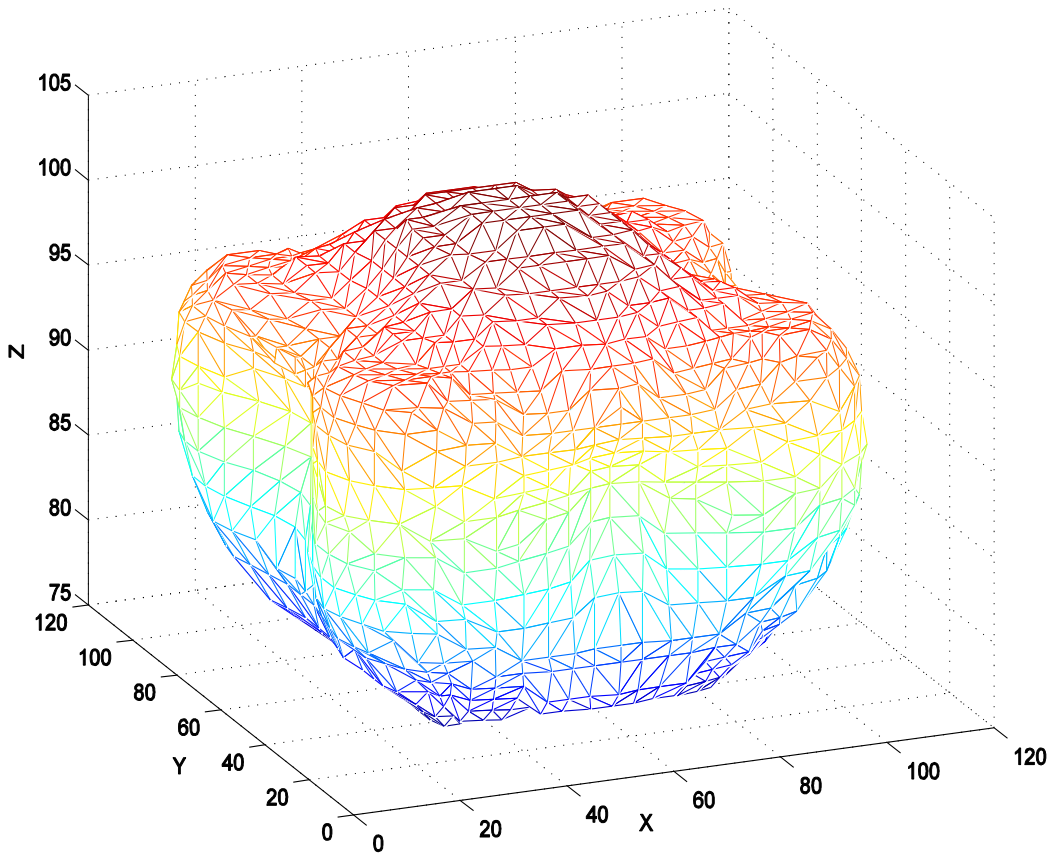

Figure 3: Original domain of a model

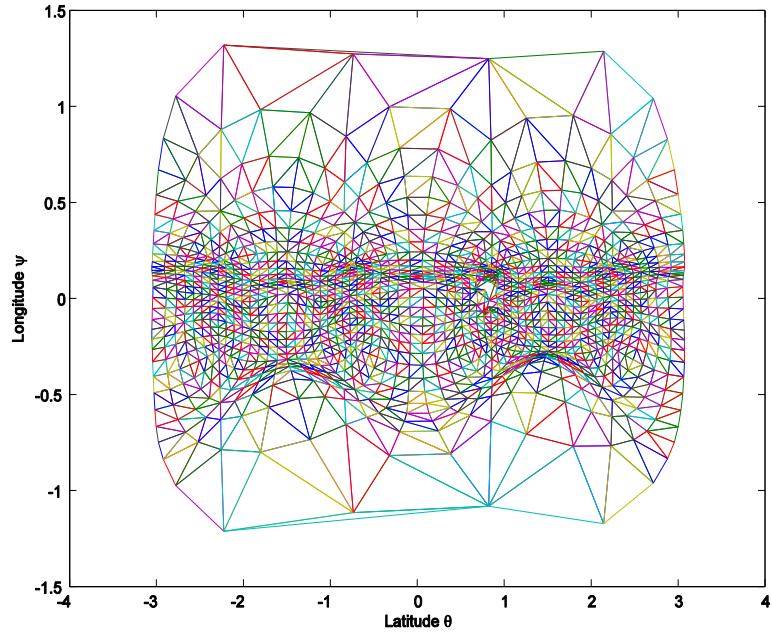

Figure 4: Atlas of a synthetic domain of a model

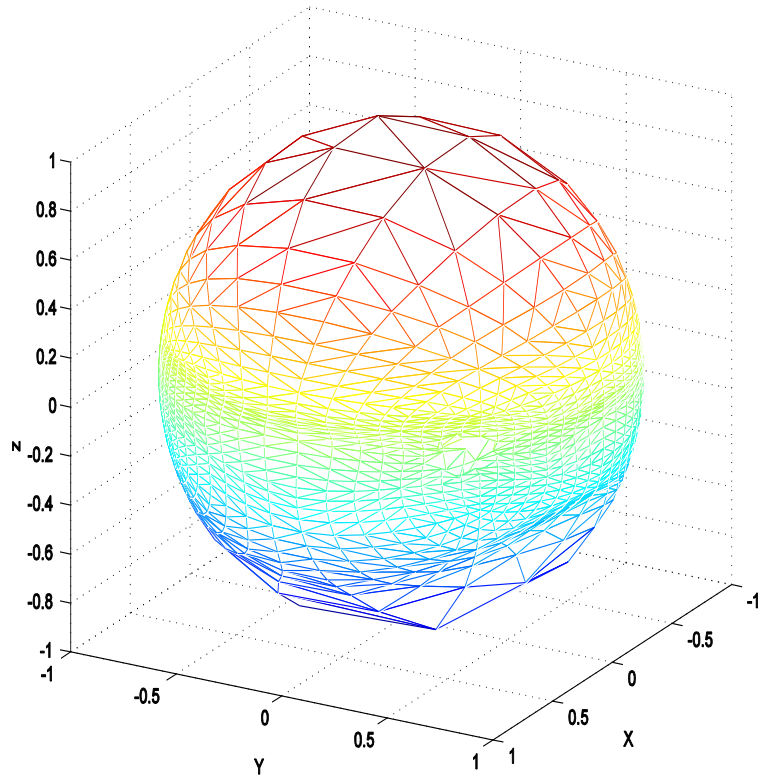

Figure 5: Synthetic domain model after mapping to a sphere

## 6.2 Case 2: Starfish

The starfish is one of the models used for testing our algorithm. It has 2745 nodes and together they make 9761 triangular faces in the domain. A value of $10^{-4}$ is used for the maximum allowed error $(\zeta)$ in potential computation. As it can be seen, the model shown in figure 6 resembles that of a starfish and the figure 7 shows the plot of the parameterized model. The dense zone of the atlas is blown up in two stages for better view and one can see that there is no bijectivity loss even in such zones. The five lobes of the starfish can be easily pointed out in the atlas.

With such a systematic parameterization method for a 3D region in place, one can exploit it for many applications like path planning. Karnik et al [10] and Voruganti et al [25] have emphasized the utility of potential field based approaches to path planning. Hence, it can be seen that volumetric parameterization of a domain through mapping is a good approach to path planning.

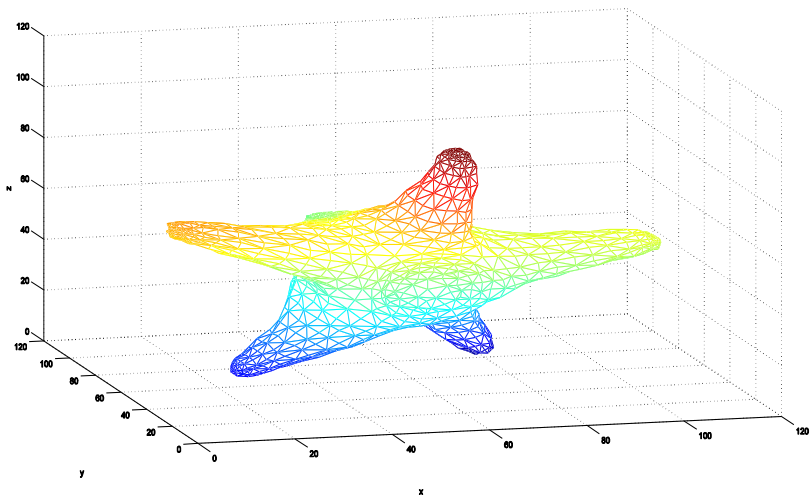

Fig.6: Original domain of a starfish.

## 6.3 Case 3: Head

The model of a human head is another interesting case that is illustrated here. The human head presents steep and frequent changes in the surface normal, near the nose and the eyes, thus making it a complex example of non-convex domain. It contains 990 nodes, which make 3206 triangular faces.
Figure 8(a) shows the target model and the corresponding parameterized atlas is shown in figure 8(b). Due to the increased complexity of the model, we face a problem in choosing an appropriate shape center. The salient features of the head, such as eyes, nose and lips can easily be pointed out in the parameterized atlas plot, which confirms that mapping is bijective.

Apart from path planning, parameterization methods are regularly used as a geometry processing methods in applications like remeshing, physical simulations, animation etc [15]. Physical simulations like stress analysis and heat transfer analysis are the applications where the existing surface parameterization methods cannot be used since the entire domain including its interior is to be parameterized. Since the proposed method provides parameterization of the entire volume at one go, it is naturally suitable for applications involving the physics of the model.

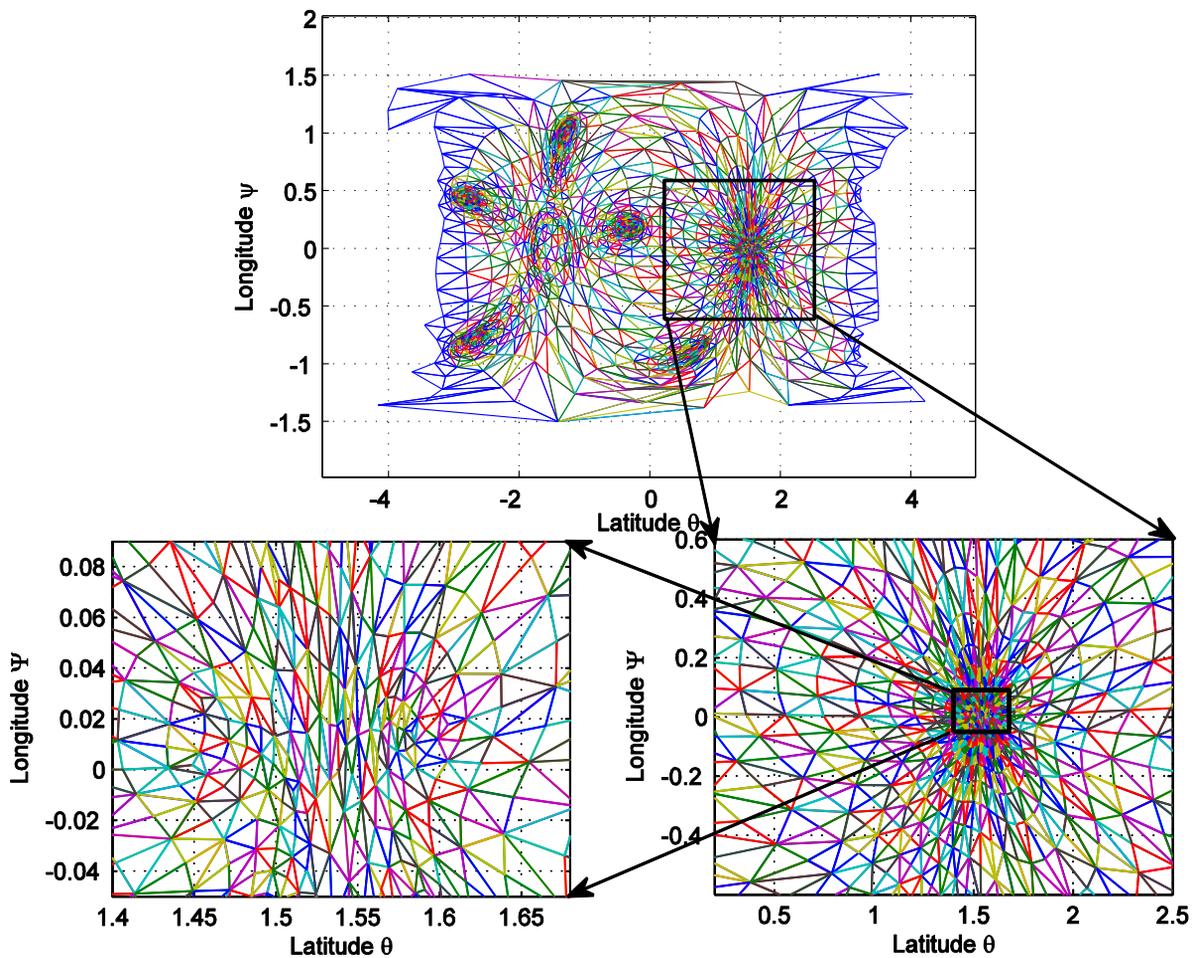

Fig.7: Atlas plot for starfish domain with two close-up views.

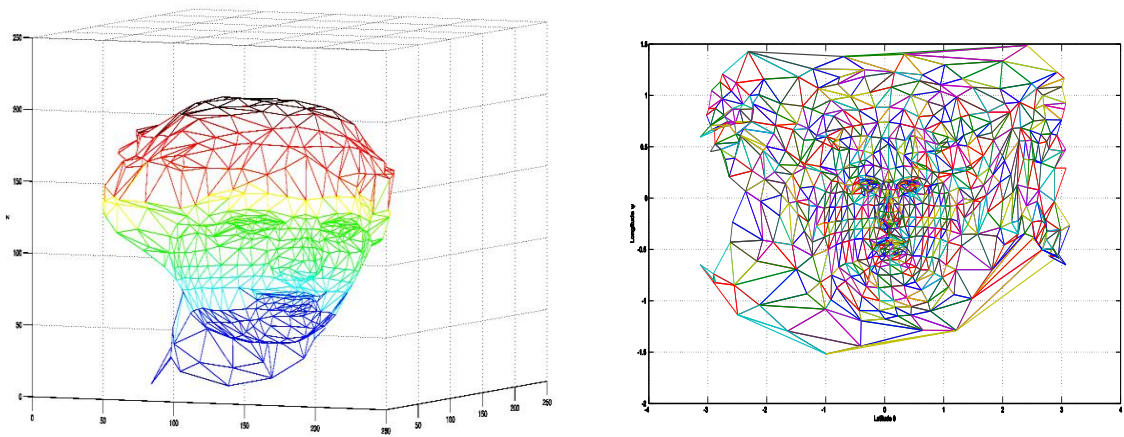

Figure 8: Domain Mapping: (a) Original domain of a human head model, (b) Atlas of the model after mapping it to a sphere.

## 6.4  Case 4: Molecule

A molecule shown in figure 9 is a fairly complex and important domain for parameterization considering the fact that this kind of parameterization has applications in many problems in computational biology. The spherical lobes of the original domain can easily be figured out in the parameterized atlas shown in figure 10.

The surface of a molecule with which it interacts with solvent or other molecules is usually obtained by rolling the probe sphere over the constituent atoms which are modeled as spheres [21]. Hence the model of a molecule looks like the boundary of the union of interiors of a set of spheres. A complete parameterization of this surface is quite useful in the study of molecular interactions. For example, in protein-protein docking problem where it is required to predict whether a given pair of proteins interact (dock), a continuous representation of protein surfaces is required for the shape matching exercise. A global parameterization scheme would also facilitate computation of curvatures and exploitation of good conformation for docking.

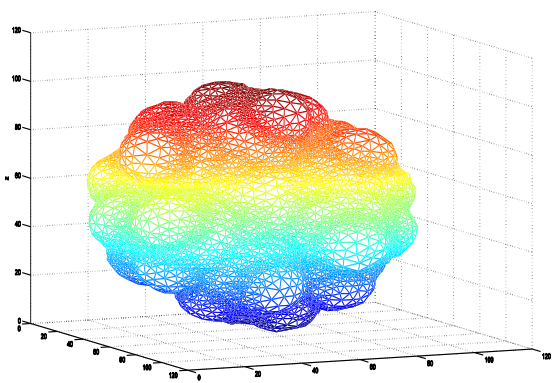

Figure 9: Original domain of a typical molecule

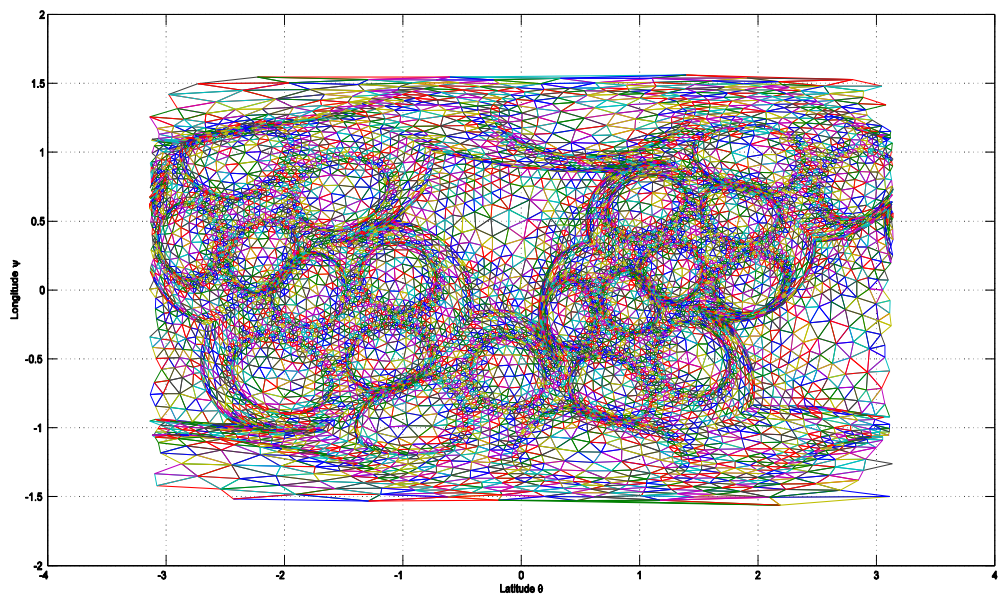

Figure 10: Atlas of the model after mapping it to a sphere.

## 7    CONCLUSIONS

In this paper, a novel approach is presented to parameterize the volume of a 3D non-convex domain using harmonic function theory. Such volumetric parameterization can be utilized for path planning and shape matching applications. Though there are various other ways to parameterize the boundary of a non-convex domain [4,19,23], only a handful of them exists for volumetric parameterization. Moreover, our approach to this problem is fairly straight forward and intuitive. Though there cannot be any loss of bijectivity, it may occur in some cases because of faults in the input data or its triangulation. Even in those cases, the loss of bijectivity will be local. A loss in bijectivity on a global scale will not occur at all because of the way potentials are computed. It is also to be noted that this method can be used as a geometry processing tool in many application problems like robot path planning, protein-protein docking, remeshing, animation etc.

There is sufficient room for future work in this area. One of the most important aspects that needs attention is to find a way to quantify the cumulative distortion of a model in mapping. Once it is quantified, one can devise a way to minimize it and obtain a better quality of mapping. One can also extend this methodology to higher dimensions. For example, in robot path planning, the dimension of the configuration space is the number joints of a robot. Usually, path planning is performed in this space. Volumetric parameterization can be used to map this higher dimensional space and plan paths.